\documentclass[12pt,preprint]{aastex}

\begin{document}

\title{Response to the Mouschovias-Tassis Comments on ``Testing Magnetic Star Formation Theory''}

\author{Richard M. Crutcher, Nicholas Hakobian}
\affil{Astronomy Department, University of Illinois, Urbana, IL 61801}
\email{crutcher@illinois.edu, nhakobi2@astro.uiuc.edu}
\and
\author{Thomas H. Troland}
\affil{Physics and Astronomy Department, University of Kentucky,
    Lexington, KY 40506}
\email{troland@pa.uky.edu}

\section{Introduction}

Our paper ``Testing Magnetic Star Formation Theory'' \citep{CHT08} (hereafter CHT) has been severely criticized by \citet{MT08} (hereafter MT). Although all of their criticisms were addressed implicitly in our paper, in the sense that we provided rationale for the observational program and the analysis of those observations, the paper did not explicitly discuss in detail the matters they claimed to be flaws in our analysis. Here we briefly do that; our brevity requires reading CHT and MT before this note.

\section{The Heart of the Matter}

The ambipolar diffusion (AD) theory {\em absolutely requires} that the mass-to-flux ratio ($M/\Phi$) increase in the central region from an initial subcritical value that supports the entire cloud against gravity. Previous measurements of $M/\Phi$ toward cloud cores have not definitively distinguished between AD and turbulence driving of star formation, e.g., \citet{TC08}. The goal of our experiment was to attempt a more definitive test. In order to carry out a quantitative and unbiased test, it is necessary to have detailed, quantitative AD models with which to compare the data. We designed our experiment to test the models that have been published by the AD theorists. Specifically, in CHT we discuss three such AD models: (1) a dimensionless parameter model \citep{CM94} for which they listed specific physical parameters for comparison with actual molecular clouds; this model with their listed physical parameters had an unevolved cloud radius of 4.3 pc; (2) one specifically computed for L1544 \citep{CB00} with an unevolved cloud radius of about 2.5 pc; and (3) a model for B1 with an unevolved cloud radius of 2.9 pc. These radii would become the radii of the ``envelopes'' surrounding the cores formed by AD, since the region of the cloud outside the cores would be ``held in place'' (an AD theorist phrase) by their subcritical magnetic fields. Moreover, the radii of the cores in these models were all $\sim 0.1$ pc. So we needed to sample the core with a filled beam of radius $\sim 0.1$ pc and the envelope with a beam radius less than $\sim 2$ pc that excluded the core. A second aspect of the numerical AD models was that the magnetic fields were smooth and regular within their unevolved cloud radii, although with an ``hourglass'' morphology strongest in the core. Although this regularity of the field was by construction in the models, the requirement that $M/\Phi$ in the clouds be subcritical meant that the magnetic fields must be strong. For the  initial field strengths and densities in the above AD models together with the non-thermal line widths in the four clouds CHT studied, the requirement for essentially ordered and regular field lines -- that magnetic energy dominates turbulent energy -- is met. Hence, these two inputs from the actual AD models guided our design of the experiment -- we needed to sample cores and envelopes on the relevant spatial scales and we could (to first order) assume that the magnetic fields in the clouds were not significantly twisted but mainly ordered. 

We chose four clouds for which there were significant detections of the OH Zeeman effect. Two of these were in the Taurus molecular cloud complex (distance $\sim 150$ pc) and two in the Perseus molecular cloud complex (distance $\sim 300$ pc). A 0.1 pc core radius would be  $\sim 1^\prime (2^\prime)$ and a 2 pc cloud (or envelope) radius would be $\sim 20^\prime (40^\prime)$ at the 300 (150) pc distance. At the OH line frequency, the primary-beam radius of the Arecibo telescope is $\sim 1.5^\prime$, so this is well matched for measurement of core properties at the densities sampled by OH. The Green Bank telescope (GBT) beam radius is $\sim 3.9^\prime$, and we pointed the GBT at positions $6^\prime$ from the Arecibo pointing position. Therefore, the GBT beams exclude the molecular core and sample the radius range $2.1^\prime$ to $9.9^\prime$, or $\sim 0.2 (0.1)$ pc to $\sim 0.9 (0.45)$ pc at the 300 (150) pc distance. This choice of beams was considered and deliberate -- set by the specifications of the AD models cited above. The sampling of the envelope had to be sufficiently far from the core to obtain a significantly different result from the core result, but {\em well} within the outer boundary of the unevolved clouds; the GBT beamsize was ideally suited to these objectives. We ``synthesized'' a toroidal beam to sample the envelopes by arithmetically adding the observations from the four GBT beams. This produced a toroidal beam, not uniformly but nonetheless reasonably well sampled around the torus -- exactly the envelope sampling called for by the AD models. 

The heart of the objections made by MT come down to whether it is appropriate to azimuthally average results of the four GBT measurements in the envelope. We strongly disagree with this heart of their argument -- that this is a ``flaw'' that leads to completely incorrect results. The issue is whether this spatial variation is what dominates the physics, or whether the azimuthally-averaged radial variation of observed parameters captures the important physics. AT theorists have argued that such small-scale structure produced perhaps by turbulence is not central to the physics of star formation; in testing the ``idealized'' AD models, we follow their lead. MT show a cartoon of the possible morphology of magnetic field lines around a single core, another cartoon with twisting and reversing field direction between four cores in a flux tube, and discuss these cartoons qualitatively. These qualitative complications give the freedom to explain virtually any observational test of the AD model; if this assertion is accepted, AD models of star formation are not testable by experiment or observation -- the essence of the scientific method. Further, many investigators have implicitly agreed that azimuthal averaging is valid -- for example, in measuring the radial profiles of cores for comparison with Bonner-Ebert profiles. If such azimuthal averaging were agreed to be a ``flawed'' approach, many of the analyzes in the literature would have to be so judged. When one spatially resolves such objects, one always finds small-scale structure, particularly as you go to observational probes of higher densities. We (along with apparently the remainder of the astronomy community aside from MT) argue that it is meaningful to use mean values and the uncertainties in those mean values to study astrophysical phenomena.

As we clearly stated, CHT tested the published, quantitative models, not ``models'' with ad hoc qualitative complications. The central point is whether or not it is valid to test quantitative  theoretical models that presumably capture the important physics, or whether it is an appropriate scientific method to add ad hoc, qualitative complications that make the theory untestable.

\section{Specific major ``flaws'' claimed by MT}

We now discuss one by one the specific ``flaws'' listed by MT.
 
\subsection{The treatment of the propagation of observational uncertainties systematically underestimates the uncertainties of the combined result}

We have ``synthesized'' a toroidal beam with parameters chosen to match the sampling scale defined by the published AD models. There are two ways to produce this synthesis. One is to fit for the magnetic field at each position and then arithmetically average the results. This is the method that MT argue to be flawed by presenting a tutorial on error analysis that obfuscates the central point. The other method is to average the Stokes $I$ and $V$ spectra from the four beams, and to infer a mean line-of-sight field ($B_{los}$) in the envelope by fitting in the standard way $dI/d\nu$ to Stokes $V$. This we have also done. As expected, since the arithmetic averaging of the four individual $B_{los}$ results is a linear process, as is the arithmetic averaging of the Stokes $I$ and $V$ spectra, the results for the mean $B_{los}$ and its uncertainty are the same. The appropriate uncertainty depends on what one is interested in. MT argue that one should compute a mean value, then base the uncertainty estimate on the differences between each of the measured values and the mean (their equation 10), taking the measurement errors into account. That does give an estimate for the range of the measured values. But what we are interested in is how accurately the mean value of $B_{los}$ in the envelope has been measured. That depends on the accuracy of the measurements. (For example, if the mean and the uncertainty in the mean of the weights of 10 people whose weights ranged from 100 to 300 pounds were desired, and the accuracy of the scales were $\pm 3$ pounds, the uncertainty in the mean weight would be $3/\sqrt{10-1} = 1$ pound. The dispersion in the weights would be much larger, the MT point.) If there are a very small number of measurements, the error estimates can best by established by the sophisticated techniques discussed by MT. But in our case, we get the uncertainty in the mean $B_{los}$ directly from the fitting of the summed Stokes I and V spectra, which is essentially the same as the uncertainty obtained by error propagation from the four individual fits. A single toroidal beam would give a $B_{los}$ and its uncertainty. That is what CHT used, since that is what is relevant for comparison with the ``idealized'' models. So the issue is not an esoteric error analysis, the issue is purely whether or not it is a valid approach to determine a mean line-of-sight field strength and a mean column density in the envelope and to use those in the analysis of the predictions of the published, quantitative AD models. We have argued above that this is a correct analysis procedure {\em to test the published models}.  Claiming by reference to cartoons that anything measurable is consistent with AD (e.g., field direction reversals in the envelope regions where the models themselves {\em require} strong magnetic fields that would prevent such twists and reversals) is not something testable by observation. Indeed, MT say that ``... observations, when reliable, would provide useful {\em input} to the ambipolar-diffusion theory'', i.e., not test the theory. What is the role in science of a theory that is not testable?
 
\subsection{Nondetections of magnetic fields are treated as if they were detections, and upper limits are not quoted as appropriate}

This is an irrelevant point. We have obtained {\em measurements} with errors for the mean line-of-sight fields in the envelopes. In using these measured values with errors to estimate ${\cal R}$ or ${\cal R^\prime}$ (the change in the mass-to-flux ratio from envelope to core), we state the difference from 1 as so many $\sigma$. For example, for L1448 we give ${\cal R^\prime} = 0.25 \pm 0.29$, which is $2.6\sigma < 1$. We could have said that a $2\sigma$ (95.4\% confidence level) upper limit on ${\cal R^\prime}$ is 0.83. What matters is how many $\sigma$ one requires in order to consider a result to be significant. We prefer to give the measurement (even if it is not a detection) and its $1\sigma$ uncertainty; readers may then consider what confidence level they require before considering a result to be significant. However, in such considerations, remember that the AD prediction is not ${\cal R^\prime} =1$, which would be for an initially exactly critical cloud, but ${\cal R^\prime} \approx 1/\mu$, where $\mu$ is the fraction by which the unevolved model cloud was subcritical. Typical AD models have $\mu \sim 1/2$, so the significance of our measured ${\cal R^\prime}$ values  with respect to the AD models should be compared not with ${\cal R^\prime} = 1$ but with ${\cal R^\prime} \sim 2$. 
 
\subsection{The theoretical simulations chosen for this comparison use as input a very different geometry of the field lines threading the cloud from the geometry of the field lines that the observations presumably reveal}

The geometry of the field lines chosen for {\em all} the theoretical simulations is a regular one, i.e., no semi-random wandering of the direction of the magnetic field, at least {\em on the scale of an individual cloud}. For two of the clouds (L1448 and B217-2), there is no detection of a field in any of the GBT beams; that is, all of the measurements are consistent with $B_{los} \approx 0$. Based on these observations, one cannot claim evidence for any different field morphology from the regular one of the theoretical simulations. For L1544 there clearly is structure in $B_{los}$, with the value at the west envelope position being about twice that at the core position, while the other three envelope position results are consistent with zero. For B1 the east and west envelope results are consistent with $B_{los} \approx 0$, while there is evidence for a north-south ridge in $B_{los}$ based on the other envelope positions. So the Zeeman data are inconsistent with the ``idealized'' AD models without any further analysis. Our analysis shows quantitatively how insistent the data are with the ``idealized'' theory.

MT state that we believe our data show field reversals. This statement is incorrect. Because many of the measured $B_{los}$ are close to zero, several GBT values for $B_{los}$ toward the same cloud have opposite sign. We stated that although nominally this would indicate a field reversal, none of these nominal reversals were statistically significant. Also, as CHT noted, a {\em net} difference in field direction between the envelope and core of $30^\circ$ (a large value!) would make only a 13\% difference in the comparison between the core and envelope $M/\Phi$.

As one goes to higher and higher angular resolution, you do not see the simple mass distribution of the ``idealized'' models; there may be multiple fragments, filaments, et cetera. The crucial question is whether the published, quantitative AD models are testable by using azimuthally averaged results -- a common technique in astrophysics. We have argued above that our choice of telescope beams was made to encompass {\em only} the radii over which the AD models say the magnetic field dominates, not the very extended ISM where the magnetic fields passing through cores may indeed ``randomly'' wander. The issue is whether it is valid to test these models, or whether observations can only serve as ``input'' to AD models taken to be correct.

\subsection{The magnetic flux threading each cloud is incorrectly calculated by taking the arithmetic average of the 
four algebraic values of B in each envelope}

MT want to take the {\em maximum} value of $B_{los}$ measured at any of the four GBT positions as the best measure of the magnetic field strength in each envelope, since field twisting in the envelopes could put the field in the plane of the sky and hence yield $B_{los} = 0$. The implication of this MT argument is that with severely twisting field lines, perhaps one GBT beam would pick up the ``correct'' $B_{los}$ to compare with the core value. But MT can accommodate whatever value we measure at any envelope position, even if it is zero at all four positions. In their view, this would just mean that the field that is largely or partly along the line of sight at the core position has just twisted to be in the plane of the sky at all envelope positions. They are saying that the AD theory is not testable by observations. 

\subsection{The possible breakup ... in the central flux tube of each cloud into ... magnetically subcritical fragments ... is ... not taken into consideration} 

To do this would require measuring the mass and flux in a 3D flux tube as it wanders around the Milky Way galaxy, which will never be possible. The complaint that we did not take this point into consideration is another way of saying the AD theory is not testable.

Moreover, the MT cartoon of four fragments in a twisted flux tube does not seem to be generally borne out by observations of dense cores. Dense cores such as the four we have studied are often part of a dense filament mapped in CO or other tracers. If such filaments were flux tubes, the magnetic field would be along the filament. This is sometimes seen, especially for lower density filaments. But the dense filaments in Taurus and Perseus (where our four cores reside) are perpendicular to the plane-of-sky field mapped by linearly polarized background starlight polarization \citep{Getal00, Getal08}, suggesting that the filaments were formed by motions along magnetic fields rather than being flux tubes.

\section{Concluding Remarks}

It is unclear to us why MT so strongly denigrate our experiment and its results. They appear not to have noticed the limitations and caveats we did discuss, some of which were added to our paper based on discussion with them. We do {\em not} state in our paper that we have proved the AD theory to be wrong. Our conclusion was that the results of our carefully planned experiment were not consistent with the ``idealized'' AD models, i.e., those that assume an initially spherical cloud with a uniform magnetic field.  MT attack a conclusion that does not exist in our paper. All of the ``flaws'' discussed by MT are not flaws, but limitations in testing any model that does not include all of the complexities of the real world. Based on their discussion, it is unclear to us what observational test of the AD theory could be realistically carried out that could provide MT anything more than observational ``inputs'' to the AD theory. We stand fully behind our experiment and our paper.

\end{document}